\documentclass[sigconf]{acmart}

\usepackage{algorithm, algorithmicx, algpseudocode, amsmath, amssymb}
\usepackage{booktabs}  
\usepackage[table]{xcolor}

\AtBeginDocument{%
  }

\setcopyright{acmlicensed}
\copyrightyear{2026}
\acmYear{2026}
\acmDOI{XXXXXXX.XXXXXXX}

\begin{document}

\title{Late Breaking Results: A Game Theoretic Approach for Optimizing Quantum Error Budget Distribution}


\author{Asif Akhtab Ronggon}
\affiliation{%
  \institution{Louisiana State University}
  \city{Baton Rouge}
  \state{Louisiana}
  \country{USA}
}
\email{arongg1@lsu.edu}

\author{Tasnuva Farheen}
\affiliation{%
  \institution{Louisiana State University}
  \city{Baton Rouge}
  \state{Louisiana}
  \country{USA}
}
\email{tfarheen@lsu.edu}







\begin{abstract}
Current fault-tolerant quantum compilers allocate error budgets uniformly during resource estimation, causing suboptimal physical resource overhead. We optimize this allocation using a potential game formulation, where Nash Equilibrium yields a Pareto-optimal distribution across logical operations, T-state distillation, and rotation synthesis. An iterated best response (IBR) algorithm converges to this equilibrium through monotonic descent of the shared cost function. Evaluation across 433 MQT benchmarks demonstrates an average reduction of 30.22\% in physical resource requirements relative to uniform baselines, with peak improvements of 97.81\% for specific circuit instances. This establishes a game-theoretic foundation for strategic error budget optimization in fault-tolerant quantum design automation.
\end{abstract}




\maketitle
\vspace{-1em}
\begin{center}
\footnotesize
\textit{\textbf{Accepted at the Design Automation Conference (DAC), Late-Breaking Results Track}.}
\end{center}

\section{Introduction}

Resource estimation for fault-tolerant quantum computing\cite{katabarwa2024early}  constitutes a critical bottleneck in the quantum design automation workflow, determining whether algorithms remain within the physical constraints of near-term hardware. As quantum applications scale beyond the noisy intermediate scale regime, the compilation process faces escalating demand to minimize space-time volume\cite{steane1998space} while still meeting the strict error-correction thresholds that fault-tolerant operation requires.

To meet these requirements, the global error budget representing the maximum tolerable error rate must be distributed among logical operations, T-state distillation, and rotation synthesis\cite{van2023using}. Recent work has demonstrated that strategic non uniform allocation can significantly reduce hardware requirements compared to uniform baselines, employing supervised machine learning to predict optimized distributions from labeled training datasets \cite{forster2025improving}. While effective, this approach requires extensive data curation, suffers from black box opacity, and provides no convergence guarantees within the compilation flow\cite{lumbreras2025interpretable,sun2020optimization}. Moreover, the model's accuracy depends entirely on the diversity of the training data risking unstable predictions for novel circuit topologies.

\textit{In this paper, we propose error budget allocation using intrinsic game theoretic structure that provides direct analytical solution without training data.} By formulating allocation as a common interest potential game, we establish that Nash Equilibrium corresponds to a Pareto optimal distribution where no component can unilaterally reduce resource cost without increasing total overhead. An IBR algorithm converges monotonically to this equilibrium, guaranteeing improvement at each step while eliminating training overhead. This analytical framework eliminates the data curation burden and provides deterministic convergence guarantees unavailable to learning based approaches. Evaluation across 433 MQT benchmark\cite{quetschlich2023mqtbench} circuits demonstrates 30.22\% average resource reduction, nearly double the improvement of state of the art learning based methods, establishing a principled foundation for strategic error budget optimization in fault tolerant quantum design automation.

\section{Game Formulation \& Methodology}

\begin{figure}[!t]
\centering
\includegraphics[width=\linewidth]{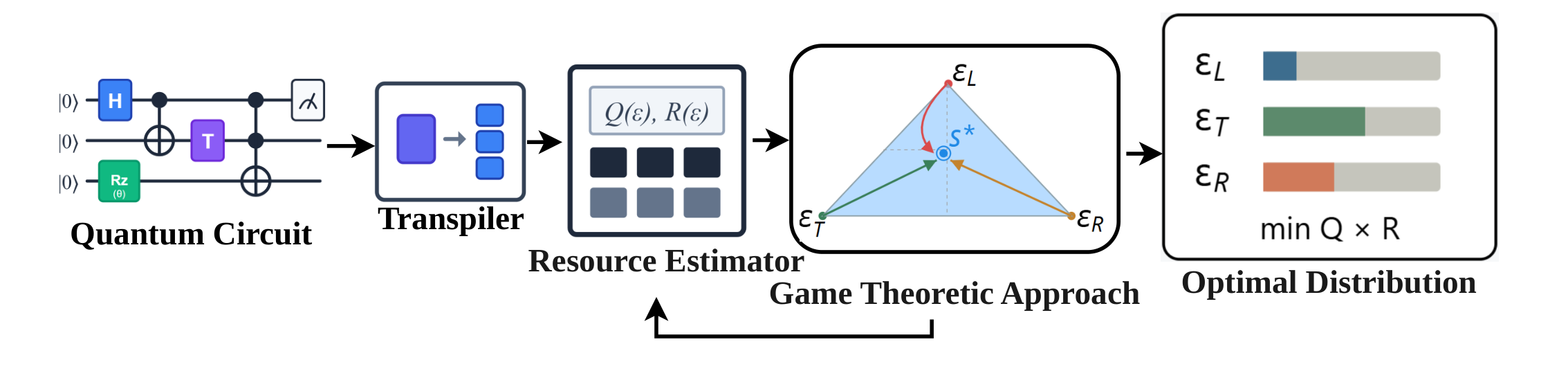}

\caption{System architecture for resource optimization.}

\label{fig:workflow}
\end{figure}

We formulate the error-budget allocation as a three-player common-interest game $\mathcal{G}=(\mathcal{N},\mathcal{S},C)$. The players $\mathcal{N}=\{L,T,R\}$ correspond to $L$, Logical error correction;$T$, T-state distillation; and $R$, Rotation synthesis. Each player $i\in\mathcal{N}$ controls a budget fraction $s_i$, with the strategy profile $\mathbf{s}=(s_L,s_T,s_R)$ constrained to the $\varepsilon$-interior of the 2-simplex $\mathcal{S}=\{\mathbf{s}\in\mathbb{R}^3_+ \mid \sum_i s_i=1,\, s_i\geq\varepsilon\}$, where $\varepsilon>0$ ensures estimator feasibility. A resource oracle maps any allocation to physical qubits $Q(\mathbf{s})$ and runtime $R(\mathbf{s})$, yielding the shared cost function $C(\mathbf{s})=Q(\mathbf{s})^w\,R(\mathbf{s})^{1-w}$ with weight $w\in(0,1)$. Since all players minimize the identical objective, $\mathcal{G}$ is an exact potential game with potential $\Phi\equiv C$; consequently, any profile minimizing $C$ constitutes a pure Nash equilibrium (NE)\cite{holt2004nash}.    Figure~\ref{fig:workflow} illustrates the compilation pipeline. The resource estimator computes physical qubit count $Q$ and runtime $R$ from the error budget partition $\boldsymbol{\varepsilon}$; the allocator minimizes their space--time product $C = Q \cdot R$.

The common-interest structure guarantees that every cost minimizing Nash equilibrium is Pareto optimal\cite{monderer1996potential}. Identical objectives prevent any unilateral deviation that improves individual payoff without increasing the shared cost $C$ , ensuring global minimizer lies on Pareto frontier. Thus, convergence to a Nash equilibrium automatically satisfies Pareto efficiency without auxiliary constraints. We compute the equilibrium through IBR. For player $i$ facing $\mathbf{s}_{-i}$, the best response minimizes $C(x_i,\mathbf{s}_{-i})$ over $x_i\in[\varepsilon,1-2\varepsilon]$. When player $i$ deviates to allocation $x_i$, the remaining budget $1-x_i$ is distributed among opponents $j$ and $k$ while preserving their fixed relative ratio $\rho = s_j/(s_j+s_k)$. This univariate optimization is solved by Brent's method~\cite{brent2013algorithms}. Cycling through players yields a potential-decreasing sequence guaranteed to converge to a pure NE. To mitigate suboptimal local equilibria, we perform $K$ random restarts $\mathbf{s}^{(0)}\sim\text{Dirichlet}(1,1,1)$ and return the minimal-cost profile, as detailed in Algorithm~\ref{alg:ibr}.

\begin{algorithm}[h] \small
\caption{Game-Theoretic Error Budget Allocation via IBR}\label{alg:ibr}
\begin{algorithmic}[1]
\Require Cost oracle $C(\cdot)$, budget allocation $\varepsilon$, weight $w$, restarts $K$, max iterations $T_{\max}$, tolerance $\delta$
\Ensure Equilibrium allocation $\mathbf{s}^{\star} \in \Delta^2$
\Statex \Comment{Cost: $C(\mathbf{s}) = Q(\mathbf{s})^{w} \cdot R(\mathbf{s})^{1-w}$, \; $Q$: physical qubits, $R$: runtime}
\State $\mathbf{s}^{\star} \gets (\tfrac{1}{3},\tfrac{1}{3},\tfrac{1}{3})$; \; $C^{\star} \gets C(\mathbf{s}^{\star})$
\For{$k = 1, \ldots, K$}
    \State $\mathbf{s}^{(0)} \sim \mathrm{Dirichlet}(1,1,1)$; \; clip to $[\epsilon, 1]$ and renormalise
    \For{$t = 0, \ldots, T_{\max}-1$}
        \State $\Delta \gets 0$
        \For{$i \in \{1,2,3\}$} \Comment{Cyclic best response}
            \State Fix ratio $\rho \gets s_j/(s_j + s_k)$ for opponents $j, k \neq i$
            \State $x_i^{\star} \gets \arg\min_{x_i \in [\epsilon,\, 1-2\epsilon]} C\bigl(\mathbf{s}(x_i, \rho)\bigr)$ \Comment{Brent's method}
            \State Update $\mathbf{s}^{(t)}$ from $x_i^{\star}$ and $\rho$; \; $\Delta \gets \max\bigl(\Delta,\, |C^{\mathrm{new}} - C^{\mathrm{old}}|\bigr)$
        \EndFor
        \If{$\Delta < \delta$} \textbf{break} \Comment{$\epsilon$-Nash condition}
        \EndIf
    \EndFor
    \If{$C(\mathbf{s}^{(t)}) < C^{\star}$} \; $\mathbf{s}^{\star} \gets \mathbf{s}^{(t)}$; \; $C^{\star} \gets C(\mathbf{s}^{(t)})$
    \EndIf
\EndFor
\State \Return $\mathbf{s}^{\star}$
\end{algorithmic}
\end{algorithm}

\section{Experimental Setup}

We evaluate our Nash equilibrium framework on 433 circuit instances from 31 circuit families in the MQT Bench suite~\cite{quetschlich2023mqtbench}, spanning state preparation, quantum algorithms, and arithmetic primitives. Circuits are transpiled using Qiskit~2.3.0 (optimization level=3) and characterized through the Azure Quantum Resource Estimator ($\varepsilon_{\text{total}}=0.1$). The allocation of error budgets is formulated as a three-player game of common-interest solved through IBR with the $\varepsilon$ -Nash threshold=$10^{-6}$, the minimum allocation= $0.005$, and the cost weight= $w=0.5$, compared to a uniform baseline $1/3$.

\section{Result}

Table~\ref{tab:comparison} establishes that equilibrium-based allocation achieves 1.94×  improvement over state-of-the-art learning-based approach (30.22\% versus 15.6\%~\cite{forster2025improving}), with peak reductions reaching 97.81\% across an expanded benchmark suite of 433 circuits spanning 31 families.

\begin{table}[H] \small
\centering

\caption{Resource improvement over uniform budget allocation.}

\label{tab:comparison}
\begin{tabular}{@{}lccc@{}}
\toprule
\textbf{Methodology} & \textbf{Samples} & \textbf{Average} & \textbf{Maximum} \\
\midrule
Forster et al.~\cite{forster2025improving} & 383 & 15.6\% & 77.7\% \\
\rowcolor{gray!15} \textbf{This work} & \textbf{433} & \textbf{30.22\%} & \textbf{97.81\%} \\
\bottomrule
\end{tabular}
\end{table}

\begin{figure}[H]
\centering
\includegraphics[width=\linewidth]{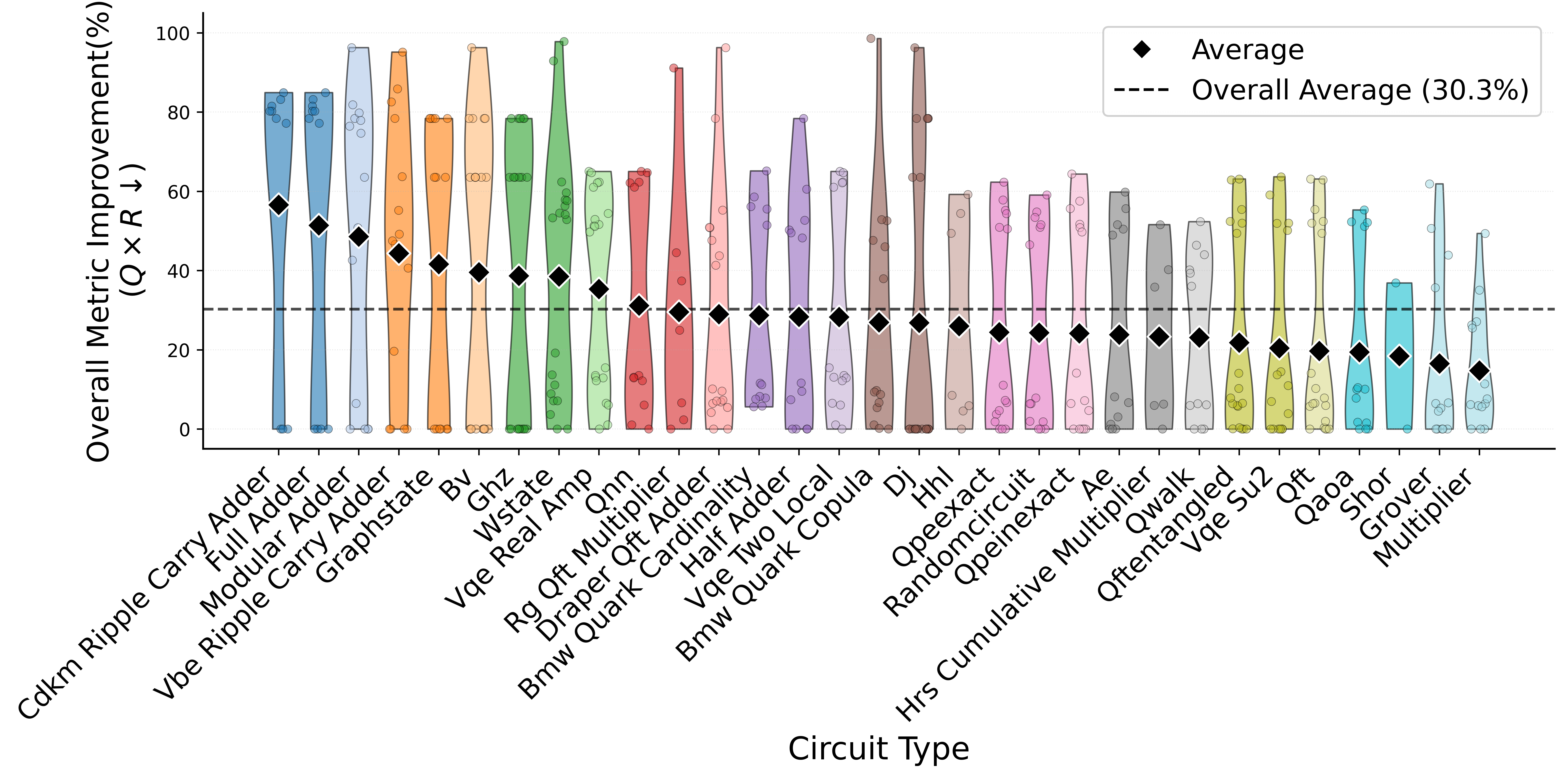}

\caption{Resource metric improvements by circuit family.}

\label{fig:improvements}
\end{figure}

Figure~\ref{fig:improvements} reveals significant heterogeneity in \textit{Overall Metric Improvement, defined as the space time volume computed as physical qubits multiplied by runtime}, relative to the uniform distribution baseline, with gains strongly correlated with circuit architectural properties. Arithmetic primitives (e.g., ripple-carry adders) and state-preparation circuits (GHZ, W-state) exhibit right-skewed distributions with extensive upper tails approaching 100\% improvement, whereas variational algorithms (QAOA, VQE) demonstrate tight variance around the mean. This dichotomy reflects structural asymmetries in error-budget sensitivity: circuits dominated by T-state distillation or high-depth logical operations present exploitable bottlenecks where aggressive redistribution yields disproportionate returns, while balanced workloads permit only marginal gains through uniform reallocation.

Critically, the game-theoretic framework captures these asymmetries without circuit-specific training or feature engineering. The monotonic convergence guarantee ensures robust performance even for outlier instances where uniform allocation incurs catastrophic overhead, establishing a principled foundation for resource optimization that generalizes across disparate circuit topologies.

\section{Conclusion}

In this work, we allocate error budgets using a natural game-theoretic formulation, achieving 30.22\% average resource reduction while providing convergence and Pareto optimality guarantees without requiring supervised learning overhead. The current methodology regarding fixed cost weighting and three-player structure presents opportunities for extension to heterogeneous architectures and additional error sources. Moreover, utilizing compiler optimization as a strategic interaction among competing resource offers a unifying framework applicable to gate scheduling, qubit mapping, and cross-layer co-design in fault-tolerant quantum systems.

\bibliographystyle{ACM-Reference-Format}
\bibliography{ref}

\end{document}